\documentclass[conference]{IEEEtran}
\IEEEoverridecommandlockouts
\usepackage{cite}
\usepackage{amsmath,amssymb,amsfonts}
\usepackage{algorithm}
\usepackage{algpseudocode}
\usepackage{float}
\usepackage{bbm}
\usepackage{graphicx}
\usepackage{textcomp}
\usepackage{xcolor}
\def\BibTeX{{\rm B\kern-.05em{\sc i\kern-.025em b}\kern-.08em
    T\kern-.1667em\lower.7ex\hbox{E}\kern-.125emX}}
    
    \usepackage{mleftright}
\mleftright
\usepackage{balance}

\usepackage{caption}
  \usepackage{comment}
    
\begin{document}

\title{Interference Reduction in Virtual Cell Optimization
}

\author{%
	  \IEEEauthorblockN{Michal Yemini}
  \IEEEauthorblockA{Princeton University\\
  	New Jersey, USA \\
  	Email: myemini@princeton.edu \thanks{This work was partially supported by the AFOSR award \#002484665 and partially supported by the NSF grant \#1824434.}}
  \and
   \IEEEauthorblockN{Elza Erkip}
  \IEEEauthorblockA{New York University\\
  	New York, USA \\
  	Email: elza@nyu.edu}
  \and
  \IEEEauthorblockN{Andrea J. Goldsmith}
  \IEEEauthorblockA{	Princeton University\\
  	New Jersey, USA \\
  	Email: goldsmith@princeton.edu}
}

\newtheorem{definition}{Definition}
\maketitle

\begin{abstract}
	Virtual cell optimization clusters cells into neighborhoods and  performs optimized resource allocation over each neighborhood.
	 In prior works we proposed resource allocation schemes to mitigate the interference caused by transmissions in the same virtual cell. 
	 This work aims at mitigating both the interference caused by the transmissions of users in the same virtual cell and the interference between transmissions in different virtual cells. We propose a resource allocation technique that reduces the number of users that cannot achieve their constant guaranteed bit rate, i.e., the ``unsatisfied users", in an uplink virtual cell system with cooperative decoding. The proposed scheme requires only the knowledge of the number of users each base station serves and relies on creating the interference graph between base stations at the edges of virtual cells.  Allocation of frequency bands to users is based on the number of users each base station would serve in a non cooperative setup. We evaluate the performance of our scheme for a mmWave system. Our
	numerical results show that our scheme decreases the number of users in the system whose rate falls below the guaranteed rate, set to $128$kbps, $256$kbps  or $512$kbps, when compared with our previously proposed optimization methods.

\end{abstract}

\section{Introduction}
Increasing demands for wireless data can only be met by increasing capacity in next-generation cellular networks.  
A prominent technique to improve network capacity is the
deployment of spectrally-efficient
small cells \cite{4623708,6768783,6171992,anpalagan_bennis_vannithamby_2015}. The importance of small cells is enhanced by the emergence of millimeter wave (mmWave) communication that exploits the significant bandwidth available at high frequency bands but suffers from high path loss, as described  in \cite{5876482,5783993,6732923,7010535,8653366}. However, the proximity of small cell base stations (BSs) to one another can cause severe interference. This interference must be managed carefully to maximize the overall network capacity. 

 One solution to mitigate the performance degradation caused by interference between BSs is BS cooperation or clustering  \cite{7839266}. 
 In this paradigm BSs fully or partially cooperate to increase the communication rates experienced by the users they serve. BS clustering and cooperation models can be categorized in many ways: A notable categorization separates BS clustering  into two groups, overlapping and non-overlapping. In overlapping  clustering,  two users can be served by two intersecting but not identical sets of BSs. 
   Overlapping clustering and BS cooperation has been extensively studied \cite{4385782,4729770,5351392,CIT-048,6786390,7248710,8110665,7470561,8755921}; of particular interest are the works \cite{4385782,4729770,5351392,CIT-048} that provide analytical guarantees on the achievable communication rates for the Wyner system model \cite{340450}. Overlapping clusters do not suffer from edge effects where users and BSs at the edges of a cluster experience  interference from the transmissions of neighboring clusters. However, this is achieved at the cost of limiting the cooperation between BSs that serve intersecting groups of users, compared with  non-overlapping clustering.  
   In  non-overlapping clustering,  two users are served by either the same set of BSs or by two disjoint sets of BSs. In this case,
 several BSs are clustered together and transmit and receive their signal cooperatively to better serve their respective users.
BS clustering can also be divided into two other main groups, namely, static clustering and dynamic clustering. Static clustering  assumes a fixed cluster of BSs. Dynamic clustering adapts to the changes in the network and overall achieves better system throughput.  This, however, entails significant overhead for channel estimation and information passing across BSs, and  requires high rate backhaul links between the BSs in the network.  While adaptive BS clustering has received much attention in the literature  \cite{5371570,4533793,6398781,8755921,6488494,6858023},  static clustering is not as well studied. The work \cite{MarschICC2011}, on the other hand, presents a method to compare clustering choices regardless of the actual user location, however, it is assumed that the clustering schemes to be compared are chosen in advance. 
The works  \cite{6530435,6707857,9343767} on static clustering are limited to systems with  non-overlapping equal size hexagonal cells.

  Our previous works
  \cite{YeminiGoldsmith1,YeminiGoldsmith2,YeminiGoldsmithJounal} develop a resource allocation optimization scheme that clusters BSs and users in the network to create entities that we name ``virtual cells'', each  comprising a set of BSs and the users they jointly serve. Our model distinguishes between two types of channel state information (CSI): the global information about the network which is used in creating clusters of BSs that we call virtual BSs, and the local instantaneous CSI. This local CSI is used to associate users with a cluster of BSs, and to allocate resources inside virtual cells. To reduce system overhead and the need for high rate backhaul links between any two BSs in the system, our scheme relies on hierarchical clustering.  Under hierarchical clustering, BSs are clustered to predefined (static) hierarchical groups, where a new cluster can be created only by merging two existing clusters. Then, the users in the BS clusters look to optimally allocate their power to maximize the cluster sum rate.
  The resource allocation techniques used in \cite{YeminiGoldsmith1,YeminiGoldsmith2,YeminiGoldsmithJounal} are successful in reducing interference in the network by jointly allocating power and channels to users in a virtual cell to mitigate the interference within that virtual cell.
   Nonetheless, the optimization in these prior works does not actively cancel  interference for edge users. Hence the rates of these edge users can be drastically reduced due to interference.  
   
   This work aims to mitigate the effect of interference \textit{between} virtual cells while limiting the requirement of additional CSI from outside each virtual cell. 
  In particular, we assume that the BSs across the network only exchange knowledge of the number of users that they serve to allocate frequency bands. 
  We propose a frequency allocation scheme that creates an interference graph that depicts interfering BSs based on their distance.
  The nodes of this graph are the BSs in the network. Two BS nodes are connected by an edge if they interfere with one another. We assume that two BSs interfere with one another if they do not belong to the same virtual BS cluster and the distance between them is smaller than a predefined threshold. Therefore, the interference graph  only depends on the choice virtual BSs and the predefined distance threshold and is static. 
  Once the interference graph is created, we allocate frequency bands to the BSs based on the number of users they serve. 
  This can be performed either by selecting a designated BS in the network to allocate frequency bands, or by a network central controller, if such exists. 
  We measure the performance of our scheme by the number of
  {\em unsatisfied users} in the system, defined as those who cannot achieve their constant guaranteed bit rate (CGBR), which is assumed to be the same across all the users in the network.
  Our numerical results demonstrate the reduction in the number of unsatisfied users for a mmWave cellular network.
  We note that our scheme does not require complex scheduling or time sharing schemes inside or outside the virtual cells to achieve the CGBR. Furthermore, our scheme is also compatible with cooperative decoding schemes with limited-capacity backhaul links that are surveyed in \cite{5594708}.
  
The remainder of the paper is organized as follows: Section \ref{sec:network_model} presents the network model and problem formulation. Section \ref{sec:virtual_cell_create} depicts the process of forming the virtual cells. Section \ref{sec:inter_reduction} proposes our interference reduction scheme. Section \ref{sec:numerical_results} presents numerical results that demonstrate the reduction in the number of unsatisfied users our scheme provides. Finally, Section \ref{sec:conclusion} concludes the paper.
 
\section{Network Model}\label{sec:network_model}
We consider  an uplink communication network that comprises a set of BSs $\mathcal{B}$, a set of users $\mathcal{U}$ and a set of frequency bands $\mathcal{K}$ each of bandwidth $W$. 
Each user $u\in\mathcal{U}$ has a power constraint of $\overline{P}_u$ dBm.
To form the  neighborhoods in which decoding is performed cooperatively, the BSs and users are clustered into virtual cells which must fulfill the following characteristics.

\subsection{Virtual Cells}\label{sec:virtual_cell_requirements}

\begin{definition}
	Let $b_1,..,b_n$ be $n$ BSs in a communication network, we call the set $\{b_1,..,b_n\}$ a virtual BS.
\end{definition}
\begin{definition}
	Let $\mathcal{B}$ be a set of BSs,  $\mathcal{U}$ be a set of users. Denote   $\mathcal{V}=\{1,\ldots,V\}$.
	For every $v$, define the sets $\mathcal{B}_v\subset \mathcal{B}$ and $\mathcal{U}_v\subset \mathcal{U}$ .
	We say that the set $\mathcal{V}$ is a proper clustering of the sets  $\mathcal{B}$ and  $\mathcal{U}$  if $\mathcal{B}_v$ and $\mathcal{U}_v$  are partitions of the sets $\mathcal{B}$ and $\mathcal{U}$, respectively. That is,
	$\bigcup_{v\in\mathcal{V}}\mathcal{B}_v = \mathcal{B}$, $\bigcup_{v\in\mathcal{U}}\mathcal{U}_v = \mathcal{U}$. Additionally,
	$\mathcal{B}_{v_1}\cap\mathcal{B}_{v_2}=\emptyset$ and $\mathcal{U}_{v_1}\cap\mathcal{U}_{v_2}=\emptyset$ for all $v_1,v_2\in\mathcal{V}$ such that $v_1\neq v_2$.
\end{definition}

\begin{definition}
	Let  $\mathcal{V}$ be a proper clustering of $\mathcal{B}$ and $\mathcal{U}$. For every $v\in\mathcal{V}$  the virtual cell  $\mathcal{C}_v$ is composed of the virtual BS $\mathcal{B}_v$ and the set of users $\mathcal{U}_v$.	
\end{definition}

This condition ensures that every BS and every user belongs to exactly one virtual cell.

Let $\mathcal{V}$ be a proper clustering of the set of BSs $\mathcal{B}$ and the set of users $\mathcal{U}$, and let   $\{\mathcal{C}_v\}_{v\in\mathcal{V}}$ be the set of virtual cells that $\mathcal{V}$ creates.
In each virtual $\mathcal{C}_v$ we assume that the BSs that compose the virtual BS $\mathcal{B}_v$ allocate their resources and decode their signal jointly.

\subsection{Uplink and CSI Models}
Let $\mathcal{B}_v$ and $\mathcal{U}_v$ be the set of BSs and the set of users that comprise the virtual cell $v$, respectively. 
In each virtual cell we consider uplink joint decoding utilizing infinite capacity links for exchanging messages between BSs within the same virtual cell. For a single frequency band scenario, this setup is equivalent to a multiple access channel  with multiple users, each with a single transmitting antenna, and  multiple receiving antennas. 
It is assumed that the channel coefficients $h_{b,u,k}: \: b\in\mathcal{B}_v, u\in\mathcal{U}_v$
between users and BSs in the virtual cell $v$ are known in advance and are used to allocate resources for transmissions within the virtual cell. Let $|G|$ be the number of elements in the set $G$.
For the sake of clarity, we label the BSs in the virtual cell $v$ by $b_1,\ldots, b_{|\mathcal{B}_v|}$.
Additionally, we denote by $p_{u,k}$ the transmission power of user $u$ in frequency band $k$ and let $\boldsymbol h_{u,k} = (h_{u,b_1,k},\ldots,h_{u,b_{|\mathcal{B}_v|},k})'$ be the channel coefficient vector between user $u$ 
to all the BSs in the virtual cell $v$. It is assumed that the covariance matrix of the interference from outside a virtual cell $v$, i.e., $\sum_{u\notin\mathcal{U}_v} p_{u,k}\boldsymbol h_{u,k} \boldsymbol h_{u,k}^{\dagger}$, is unknown during the resource allocation stage.
Nonetheless, this covariance matrix is considered to be known during the joint decoding stage. In practice it can be estimated using pilot symbols at the beginning of a transmission.

\subsection{Power Allocation in Virtual Cells}\label{subsection:uplink_joint_decoding_problem}
Denote by $x_{u,k}$ the signal of user $u$ in frequency band $k$. Additionally, denote by $\tilde{y}_{b,k}$  the received signal at BS $b$ in frequency band $k\in\mathcal{K}$, when ignoring interference from outside the virtual cell BS $b$ belongs to.
To limit interference between virtual cells, for each user $u$  we allocate a set of frequency bands $\mathcal{K}_u\subset\mathcal{K}$ that can be used for transmission, that is $x_{u,k}=0$ for all $k\notin \mathcal{K}_u$. Additionally, for each BS $b$ we allocate a set of frequency bands $K_b\subset\mathcal{K}$ that are used for receiving signals, i.e., $\tilde{y}_{b,k}=0$ for all $k\notin \mathcal{K}_b$. Since BS $b$ may suffer high interference in the bands $\mathcal{K}\setminus\mathcal{K}_b$ from neighboring virtual cells, in the power allocation stage we consider for each BS $b$ only received signals in frequency bands  $k\in\mathcal{K}_b$. 
In the decoding stage we estimate the covariance matrix of the interference from outside the virtual cell using pilot symbols. Therefore, in the decoding stage BS $b$ can use its received signals from all frequency bands. 
The received signal at BS $b\in \mathcal{B}_v$, ignoring the interference from other virtual cells, in frequency band $k\in\mathcal{K}_b$ is
\[
\tilde{y}_{b,k} =\sum_{u\in\mathcal{U}_v}h_{u,b,k} x_{u,k}+n_{b,k}
,\]
where $h_{u,b,k}$ is the channel coefficient from user $u$ in $v$ to the BS $b$ in $v$ over frequency band $k$, and $n_{b,k}$ is a white Gaussian noise at BS $b$ over frequency band $k$. 
Denote by $\boldsymbol {\tilde{y}_{v,k}}$ the column vector of the received signals $\tilde{y}_{b,k}$ at all BSs $b$ in the virtual cell $v$ such that $k\in\mathcal{K}_b$. That is,
\[\boldsymbol {\tilde{y}_{v,k}}\triangleq(\tilde{y}_{b,k})_{b:b\in\mathcal{B}_v,k\in\mathcal{K}_b}',\]
where $(\cdot)'$ denotes the transpose operator.
Additionally, denote the column vectors 
$\boldsymbol{\tilde{h}}_{u,k} = (h_{u,b,k})_{b:b\in\mathcal{B}_v,k\in\mathcal{K}_b}'$ and
$\boldsymbol{\tilde{n}}_{u,k} = (n_{b,k})_{b:b\in\mathcal{B}_v,k\in\mathcal{K}_b}'$.
 Then 
\[
\boldsymbol{\tilde y_{v,k}} = \sum_{u\in\mathcal{U}_v} \boldsymbol{\tilde{h}}_{u,k} x_{u,k}+\boldsymbol{\tilde{n}_{v,k}}.\]
Denote $\boldsymbol {\tilde{N}_{v,k}} = \text{cov}(\boldsymbol{\tilde n_{v,k}})$. The sum capacity of the uplink, ignoring interference outside the virtual cell,  in the virtual cell $v$ is then:
\begin{flalign}\label{eq:uplink_problem_clean}
\max &\sum_{k\in\mathcal{K}}W\log_2\left|\boldsymbol I+\sum_{u\in\mathcal{U}_v}p_{u,k}\boldsymbol{\tilde{h}}_{u,k} \boldsymbol{\tilde{h}}_{u,k}^{\dagger}\boldsymbol{\tilde{N}_{v,k}}^{-1}\right|\nonumber\\
\text{s.t.: } & \sum_{k\in\mathcal{K}_u} p_{u,k}\leq \overline{P}_u,\quad \forall u\in\mathcal{U}_v \nonumber\\
&\hspace{0.1cm} p_{u,k}\geq 0,\quad \forall u\in\mathcal{U}_v, k\in\mathcal{K}_u,\nonumber\\
&\hspace{0.1cm} p_{u,k}= 0,\quad \forall u\in\mathcal{U}_v, k\notin\mathcal{K}_u,
\end{flalign}
where  $\boldsymbol{\tilde{h}}_{u,k}^{\dagger}$ is the conjugate transpose of  $\boldsymbol{\tilde{h}}_{u,k}$ and $|\boldsymbol A|$ denotes the determinant of the matrix $\boldsymbol A$.  We assume that the noise covariance matrices $\boldsymbol {\tilde{N}_{v,k}}$ are invertible for all $k$ and thus also positive definite.

We note that problem (\ref{eq:uplink_problem_clean}) ignores interference that is caused by transmissions outside the virtual cell, instead it only considers interference that is caused by transmissions in the virtual cell. As  the number of virtual cells decreases, interference from inside the virtual cell   dominants the one caused by transmissions outside the virtual cell.  However, when the virtual cells are small, or in the edges of virtual cells,  BS $b$ can suffer significant
 interference in  frequency bands $\mathcal{K}\setminus\mathcal{K}_b = \{k\in\mathcal{K}:k\notin\mathcal{K}_b\}$
 caused by users in neighboring virtual cells that transmit over these frequency bands. Therefore, to  reduce the interference at BSs at the edges of the virtual cell, we limit the receiving frequency bands at edge BS $b$ in the power allocation stage to the set $\mathcal{K}_b$. Additionally, to limit interference to other virtual cells, we limit the transmission of a user $u$ at the edges of the virtual cells to the set $\mathcal{K}_u$.

The optimal solution of the convex problem \eqref{eq:uplink_problem_clean} can be found iteratively \cite{Bertsekas/99,1262622}. For completeness of presentation, we next describe the optimal iterative water-filling solution for the multi-band scenario:  

Denote  
\[\boldsymbol\Sigma_{u,k} = \boldsymbol{\tilde{N}_{v,k}}+\sum_{\substack{\tilde{u}\neq u,\\ \tilde{u}\in\mathcal{U}_v}}p_{\tilde{u},k}\boldsymbol{\tilde{h}}_{\tilde{u},k}\boldsymbol{\tilde{h}}_{\tilde{u},k}^{\dagger}.\]
Then we set 
\[p_{u,k} =  \left(\frac{W}{\lambda}-\frac{1}{\boldsymbol{\tilde{h}}_{u,k}^{\dagger}\boldsymbol\Sigma_{u,k}^{-1}\boldsymbol{\tilde{h}}_{u,k}}\right)^+\cdot\mathbbm{1}_{\{k\in\mathcal{K}_u\}}\]
where $\lambda$ is chosen such that $\sum_{k\in\mathcal{K}}p_{u,k} = \overline{P}_{u}\cdot\mathbbm{1}_{\{\mathcal{K}_u\neq\emptyset\}}$.
We iterate cyclicly between all users in $\mathcal{U}_v$ until the convergence of the power allocation of all users in $\mathcal{U}_v$.

\subsection{Virtual Cell Capacity and Unsatisfied Users}

The received signal at BS $b\in \mathcal{B}_v$,  in frequency band $k$ is
\begin{flalign}\label{eq:received_signal_actual}
y_{b,k} = \sum_{u\in\mathcal{U}_v}h_{u,b,k} x_{u,k} + \sum_{u\not\in\mathcal{U}_v}h_{u,b,k} x_{u,k} +n_{b,k}.
\end{flalign}
Note that the  term $\sum_{u\not\in\mathcal{U}_v}h_{u,b,k} x_{u,k}$ in \eqref{eq:received_signal_actual} is the interference at BS $b$ at frequency band $k$ from outside the virtual cell that BS $b$ belongs to. 

Recall that $\boldsymbol h_{u,k} = (h_{u,b_1,k},\ldots,h_{u,b_{|\mathcal{B}_v|},k})'$ is the channel coefficient vector between user $u$ to all the BSs in the virtual cell $v$. Then the received signal vectors at the BSs in $v$ are
\[
\boldsymbol y_{v,k} = \sum_{u\in\mathcal{U}_v}\boldsymbol h_{u,k} x_{u,k}+\sum_{u\notin\mathcal{U}_v}\boldsymbol h_{u,k} x_{u,k}+\boldsymbol n_{v,k}
,\]
where 
$\boldsymbol n_{v,k}=(n_{b_1,k},\ldots,n_{b_{|\mathcal{B}_v|,k}})$ is a white noise vector at the BSs.  Let $\boldsymbol {N_{v,k}} = \text{cov}(\boldsymbol n_{v,k})$ and
let $D_k(u)$ be a chosen decoding order of users at virtual cell $v$ in frequency band $k$. That is, the user $u$ such that $D_k(u)=1$ is decoded first, then user with  $D_k(u)=2$, and so on. 
Denote by 
\begin{flalign*}\boldsymbol{J}_{v,u,k}=\boldsymbol{N}_{v,k}\hspace{0.1cm}+\hspace{-0.3cm}\sum_{\substack{\tilde{u}\in\mathcal{U}_v:\\D(u)<D(\tilde{u})}}\hspace{-0.25cm}
p_{\tilde{u},k}\boldsymbol h_{\tilde{u},k} \boldsymbol h_{\tilde{u},k}^{\dagger}+\sum_{\tilde{u}\notin\mathcal{U}_v}p_{\tilde{u},k}\boldsymbol h_{\tilde{u},k} \boldsymbol h_{\tilde{u},k}^{\dagger}
\end{flalign*} 
the noise plus interference experienced by BSs in the virtual cell $v$ in frequency band $k$ when decoding the signal of user $u$. The  capacity of user $u$ in the virtual cell $v$ and frequency band $k$ is:
\begin{flalign}\label{eq:uplink_problem_with_int_k}
R(u,k)=W\log_2\left|\boldsymbol I+p_{u,k}\boldsymbol h_{u,k} \boldsymbol h_{u,k}^{\dagger}\boldsymbol{J}_{v,u,k}^{-1}\right|,
\end{flalign}
and the overall capacity of user $u$ in the virtual cell  $v$  is then:
\begin{flalign}\label{eq:uplink_problem_with_int}
R(u)&=\sum_{k\in\mathcal{K}}R(u,k)\nonumber\\
&=\sum_{k\in\mathcal{K}}W\log_2\left|\boldsymbol I+p_{u,k}\boldsymbol h_{u,k} \boldsymbol h_{u,k}^{\dagger}\boldsymbol{J}_{v,u,k}^{-1}\right|.
\end{flalign}

 The sum capacity of the uplink in the virtual cell is then $\sum_{u\in\mathcal{U}_v}R(u)$,
note that this sum capacity is not affected by the decoding order of the users.

Let $r_{GBR}$ be the guaranteed bit rate of each user. The number of unsatisfied users in the systems is then given by
\begin{flalign}\label{eq:num_unsatisfied}
\sum_{v\in\mathcal{V}}\sum_{u\in\mathcal{V}}\mathbbm{1}_{\{R(u)<r_{GBR}\}}.
\end{flalign}
Our work aims at minimizing the number of unsatisfied user in the network for our static clustering algorithm presented in Section \ref{sec:virtual_cell_create}, using low rate side information. Specifically, we use the number of users that choose each BS as its most desirable BS in terms of signal-to-noise ratio (SNR).
We assume that this is the only user dependent side information available for allocating frequency bands to BSs in virtual cells.

\section{Forming the Virtual Cells}\label{sec:virtual_cell_create}
This section presents the clustering method, proposed in \cite{YeminiGoldsmith1,YeminiGoldsmith2,YeminiGoldsmithJounal}, that creates the virtual cells within which the resource allocation scheme presented in this paper operates.
\subsection{Static Hierarchical Base-Station Clustering}\label{sec:BS_clustering}

Let $d:\mathbb{R}^2\times\mathbb{R}^2\rightarrow\mathbb{R}$ be the Euclidean distance function.  
\begin{definition}[Radius of a set around point]
	Let $S$ be a set of points in $\mathbb{R}^2$, the radius of $S$ around $s_i \in S$  is defined as $r(s_i,S)=\max_{s_j\in S}\:d(s_i,s_j)$.
\end{definition}
\begin{definition}[Minimax radius]
	Let $S$ be a set of points in $\mathbb{R}^2$, the minimax radius of $S$ is defined as $r(S) = \min_{s_i\in S}\: r(s_i,S)$.
\end{definition}
\begin{definition}[Minimax linkage]
	The minimax linkage between two sets of points $S_1$ and $S_2$ in $\mathbb{R}^l$ is defined as $d(S_1,S_2) = r(S_1\cup S_2)$.	
\end{definition}

Algorithm \ref{algo:hierarchical_clustering} clusters BS using the  hierarchical clustering algorithm using the minimax linkage criterion \cite{BienTibshirani2011} with maximal size constraint. 

Let $\mathcal{S}=\{s_1,\ldots,s_{|\mathcal{B}|}\}$ be the set of locations of the BSs in $\mathcal{B}$ and let $\overline{s}_m$ be the maximal BSs in a cluster for $m$ system clusters. Additionally, let $B_m=\{\mathcal{B}_{m,1},\ldots,\mathcal{B}_{m,m}\}$ be the  virtual clustering of BSs  for each number of clusters $m$, where $\mathcal{B}_{m,1},\ldots,\mathcal{B}_{m,m}$ are $m$ virtual BSs.
We use Algorithm \ref{algo:hierarchical_clustering} with the inputs $\mathcal{S}$ and $\overline{s}$ to create the  virtual clustering  $B_m$ for each number of clusters $m$. As discussed in \cite{YeminiGoldsmith1,YeminiGoldsmith2,YeminiGoldsmithJounal},
since interference increases on average as distance decreases, Algorithm \ref{algo:hierarchical_clustering}  merges, at each stage, two clusters to create a new one in which the minimal interference between BSs is maximized on average. This interference is then mitigated in the resource allocation stage.
Numerical results in \cite{YeminiGoldsmith1,YeminiGoldsmith2,YeminiGoldsmithJounal} show
that this clustering scheme
 yields a higher system sum rate than that of the K-means and spectral clustering algorithms. Additionally, this method enjoys a key property that both the K-means clustering and the spectral clustering lack, namely, the number of clusters can be changed without disassembling all the clusters in the networks.  This simplified the need for providing dedicated backhaul links between all BSs in the networks, and only requires backhaul links between BSs that may be merged if the system reduces its number of virtual cells.

\begin{algorithm}
	\caption{}\label{algo:hierarchical_clustering}
	\begin{algorithmic}[1]
		\State Input: $\mathcal{S}=\{s_1,\ldots,s_{|\mathcal{B}|}\}$, $\overline{s} = \{\overline{s}_1,\ldots,\overline{s}_{|\mathcal{B}|}\}$;
		\State Set $S_{|\mathcal{B}|} = \left\{\{s_1\},\dots,\{s_{|\mathcal{B}|}\}\right\}$; 
		\State Set $B_{|\mathcal{B}|} = \left\{\{b_1\},\dots,\{b_{|\mathcal{B}|}\}\right\}$; 
		\State Set $d(\{s_i\},\{s_j\})=d(s_i,s_j),\:\forall s_i,s_j\in S$;
		\For {$m = |\mathcal{B}|-1,\ldots,1$}
		\State Find $(v_1,v_2) = \arg\min_{\substack{G,H\in S_m: G\neq H,\\ |G|+|H|\leq \overline{s}_m}} d(G,H)$;\vspace{0.2cm}
		\State  Update $S_{m} = S_{m+1} \bigcup \{S_{v_1}\cup S_{v_2}\} \setminus \{S_{v_1},S_{v_2}\}$;
		\State  Update $B_{m} = B_{m+1} \bigcup \{B_{v_1}\cup B_{v_2}\} \setminus \{B_{v_1},B_{v_2}\}$;
		\State Calculate $d(S_1\cup S_2,G)$ for all $G\in S_m$;
		\EndFor
	\end{algorithmic}
\end{algorithm}

\subsection{Users' Affiliation with Clusters}\label{sec_user_affil}
We consider the ``best channel" user affiliation rule in which each user is affiliated with the BS to which it has the highest channel instantaneous coefficient in absolute value.  Then each user is associated with the virtual BS that its affiliated BS is part of.
This way every virtual BS and it associated users compose a virtual cell. This creates a proper clustering.

\section{Virtual Cell Interference Reduction}\label{sec:inter_reduction}
This section presents the dynamic frequency allocation scheme we propose for limiting interference at the edges of virtual cells. This scheme comprises two parts, the first  creates a BS interference graph based on the BS locations. The second part allocates sets of frequency bands to the BSs, based on the number of users that choose each BS as its most desirable BS in terms SNR. We then propose a decoding order to improve the minimal rates experienced by users in virtual cells. 
\subsection{BS coloring}
   We create the BS interference graph, for a given choice of BS clustering,  as follows. First, the set of vertices in the graph comprises the set of BSs $\mathcal{B}$. Let $\gamma_d$ be a threshold distance between BSs\footnote{Note that the choice of the threshold $\gamma_d$  does not affect the number of virtual cells.}. We add an edge $\{b_1,b_2\}\in\mathcal{B}^2$ if all of the following conditions hold
        \begin{enumerate}
                \item the BSs $b_1$ and $b_1$ do not belong to the same virtual cell. 
                \item the distance between BS $b_1$ and BS $b_2$ is less than $\gamma_d$, i.e., $d(b_1,b_2)<\gamma_d$. 
                \item the number of users that choose BS $b_1$ as their ``best BS" or the number of users that choose BS $b_2$ as their ``best BS" is strictly positive. 
        \end{enumerate}  
        
        Once we establish the interference graph we use a graph coloring algorithm to divide the BSs into non-interfering groups of BSs, that is, the BSs in each group do not interfere with one another. Possible candidates for graph coloring algorithms are the Recursive Largest First (RLF) \cite{RLF} or  its variations \cite{Adegbindin2014ANE}. 

\subsection{Dynamic Frequency Allocation}
\subsubsection{Dynamic BS Frequency Association for Power Allocation}
Once the interference graph is created and we partition the BSs into non-interfering groups, we allocate frequency bands to these groups proportionately to the number of users each non-interfering group serves as follows.

Suppose that that there are $\kappa$  non-interfering groups of BSs,\footnote{We assume that $|\mathcal{K}|\geq \kappa$, otherwise we reduce the threshold $\gamma_d$ to fulfill this condition.} we denote them by $\mathcal{I}_1,\ldots,\mathcal{I}_{\kappa}$.
and let $n_{\ell}$ be the number of users that choose one of the BSs in $\mathcal{I}_{\ell}$ as its ``best BS" based on the ``best channel" affiliation rule. Let $f_\ell$ be the number of frequency bands each group $\mathcal{I}_{\ell}$ receives. Denote 
\[\tilde{f}_{\ell} = |\mathcal{K}|\frac{n_{\ell}}{\sum_{i=1}^{\kappa}n_i}\quad \text{for every } \ell\in\{1,\ldots,\kappa\}.\] 
If the number of frequency bands for each group is an integer then we set $f_{\ell}=\tilde{f}_{\ell}$ for every group $\ell$. Otherwise, we allocate the number of frequency bands as follows. 

Let 
\[\delta_f = \sum_{\ell=1}^{\kappa}\lceil\tilde{f}_\ell\rceil-|\mathcal{K}|,\] and let  $r_{\ell} = \lceil\tilde{f}_{\ell}\rceil-\tilde{f}_{\ell}$. Order the sequence of remainders $(r_1,\ldots,r_{\kappa})$ from the highest to the lowest and denote this ordering by the vector $\boldsymbol{O}$.
Then the number of  frequencies associated with group $\ell$  is 
\[f_{\ell} = \lceil\tilde{f}_{\ell}\rceil-\tilde{r}_{\ell}\]
where 
$\tilde{r}_{\ell}=1$ if the index of $\ell$ is in the last $\delta_f$ indices in
$\boldsymbol{O}$ such that $\lceil\tilde{f}_{\ell}\rceil>1$, otherwise $\tilde{r}_{\ell}=0$.

Finally, for every BS $b$ in  $I_{\ell}$ we allocate the frequencies 
\[\mathcal{K}_b = \left\{\sum_{i<\ell}f_i+1,\ldots,\sum_{i<\ell}f_i+f_{\ell}\right\},\] 
where $f_0=0$.

\subsubsection{Dynamic User Frequency Allocation}
For every user $u$ that is associated with it ``best BS'' $b$ we set 
\[\mathcal{K}_u = \mathcal{K}\setminus \bigcup_{\tilde{b}\in\mathcal{\tilde{B}}}\mathcal{K}_{\tilde{b}}\]
where $\mathcal{\tilde{B}}$ is the set of all BSs $\tilde{b}$ such that 
\begin{enumerate}
\item  $b$ and $\tilde{b}$ are not in the same virtual cell, and
\item $d(b,\tilde{b})<\gamma_d$, and 
\item $d(u,\tilde{b})<\gamma_d$.
\end{enumerate}

\section{Decoding Order in Virtual Cells}
To minimize the overall number of unsatisfied users \eqref{eq:num_unsatisfied}, the optimal decoding order in each virtual cell minimizes the  number of unsatisfied users in it. This problem is complex and cannot be solved efficiently. To that end, we propose  a suboptimal and greedy algorithm, presented in  Algorithm \ref{algo:decoding_order},
to decode the signals of the users in each frequency band. This procedure decides on the decoding order serially over the frequency bands and aims at maximizing the minimal rate experienced by users at each stage greedily. 

\begin{algorithm}[H]
	\caption{Decoding order in a virtual cell}
	\begin{algorithmic}[1]
		\State Inputs: $\mathcal{U}_v$,$\mathcal{B}_v$,
		$\mathcal{K}=\{k_1,\ldots,k_{|\mathcal{K}|}\}$;\vspace{0.15cm}
		\State Inputs: $\boldsymbol{N}_{v,k}+\sum_{\tilde{u}\notin\mathcal{U}_v }p_{\tilde{u},k}\boldsymbol h_{\tilde{u},k} \boldsymbol h_{\tilde{u},k}^{\dagger}\quad \forall u\in\mathcal{U}_v, k\in\mathcal{K}$;\vspace{0.15cm}
		\State Inputs: $p_{u,k}\boldsymbol h_{u,k} \boldsymbol h_{u,k}^{\dagger}\quad \forall u\in\mathcal{U}_v, k\in\mathcal{K}$;\vspace{0.15cm}
		\State Set  $R(u,k)=0$ for every $u\in\mathcal{U}_v$ and $k\in\mathcal{K}$;
		\State For each $u\in\mathcal{U}_v$ and $k\in\mathcal{K}$ calculate 
		\begin{flalign*}
		&g(u,k) =\\
		&\log_2\left|\boldsymbol I+p_{u,k}\boldsymbol h_{u,k} \boldsymbol h_{u,k}^{\dagger}\left[\boldsymbol{N}_{v,k}+\sum_{\tilde{u}\neq u}p_{\tilde{u},k}\boldsymbol h_{\tilde{u},k} \boldsymbol h_{\tilde{u},k}^{\dagger}\right]^{-1}\right|;
		\end{flalign*}
		\State Decode the signals in frequency band $k_1$ in the descending order of $g(u,k_1)$, i.e., from strongest user to the weakest, and save their respective rates, calculated by \eqref{eq:uplink_problem_with_int_k}, in  $R(u,k_1),\:u\in\mathcal{U}_v$;
		\For {$i = 2,\ldots,|\mathcal{K}|$}
      \State Set \[\tilde{g}(u,k_i) = \sum_{j<i}R(u,k_j)+g(u,k_i);\]
      	\State Decode the signals in frequency band $k_i$ in the descending order of $\tilde{g}(u,k_i)$, and using  \eqref{eq:uplink_problem_with_int_k} calculate and save the rate achieved by each user $u\in\mathcal{U}_v$ over frequency band $k_i$ in $R(u,k_i)$;
		\EndFor
	\end{algorithmic}
\label{algo:decoding_order}
\end{algorithm}

\section{Numerical Results}\label{sec:numerical_results}
This section presents numerical results that depict the performance of our edge-user interference reduction technique for virtual cells in a mmwave cellular network. We set the following parameters for the simulation: the network is comprised of $20$ single antenna BSs and $200$ single antenna users which are uniformly located in a square of side $400$ meters.  There are $24$ equal bandwidth frequency bands, the total bandwidth is $5$MHz, the carrier frequency is set to $28$GHz. The noise power received by each BS is $-174$ dBm/Hz, and the maximal power constraint for each user is $23$ dBm.  
The channel coefficients $h_{u,b,k}$ between the single antenna transmitters and receivers are generated according to \cite{6834753} where $h_{u,b,k}=0$ when a blockage occurs. Additionally, to present the full effect of reducing the number of virtual cells serially, the maximal number of BSs in a virtual cells were chosen according to the number of leaves at each depth of a binary tree with $20$ leaves. Specifically, for number of virtual cells equal or less than $10$ the maximal number of BSs in a virtual cell was two, for number of virtual cells equal or less than five the maximal number of BSs in a virtual cell was four,  and so on. 
We averaged the results over 500 system realizations, in each we generated randomly the locations of the BSs, users and channel coefficients.

\begin{figure}
    \centering
    \includegraphics[scale=0.66]{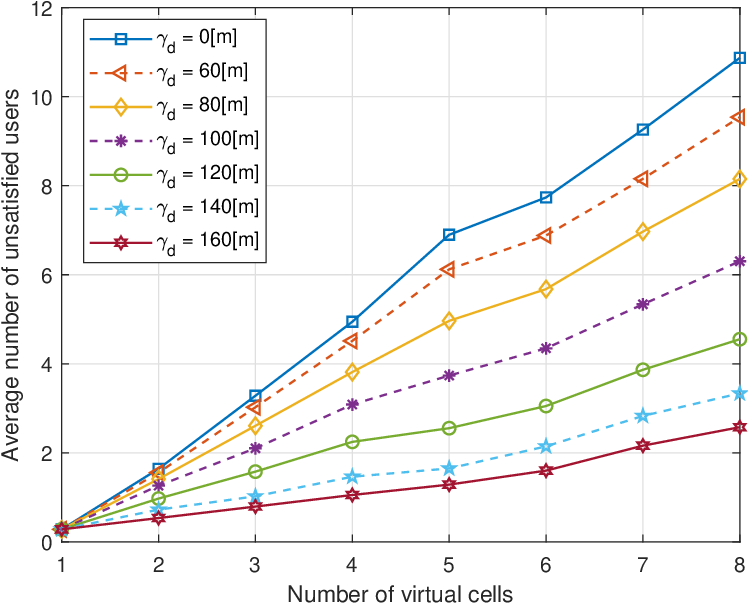}
    \caption{Number of unsatisfied users for a CGBR of 128kbps.}
    \label{fig:my_label128}
\end{figure}

\begin{figure}
    \centering
    \includegraphics[scale=0.66]{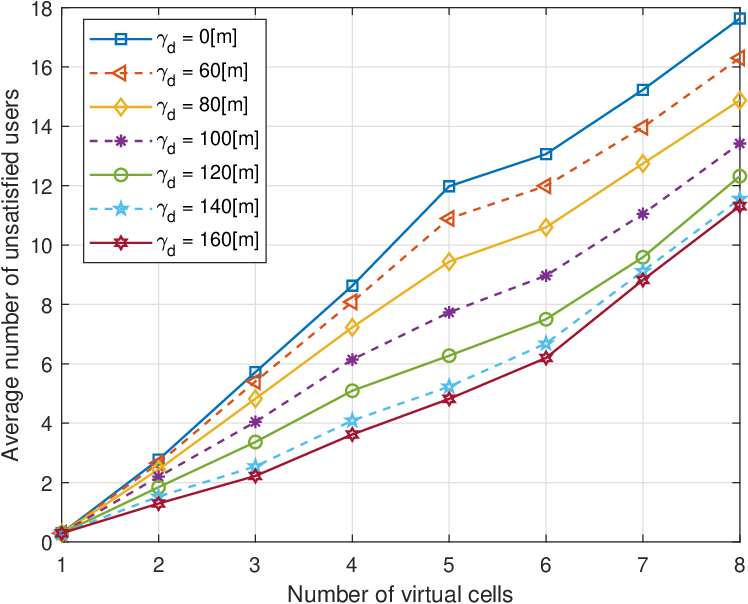}
    \caption{Number of unsatisfied users for a CGBR of 256kbps.}
    \label{fig:my_label256}
\end{figure}

\begin{figure}
    \centering
    \includegraphics[scale=0.66]{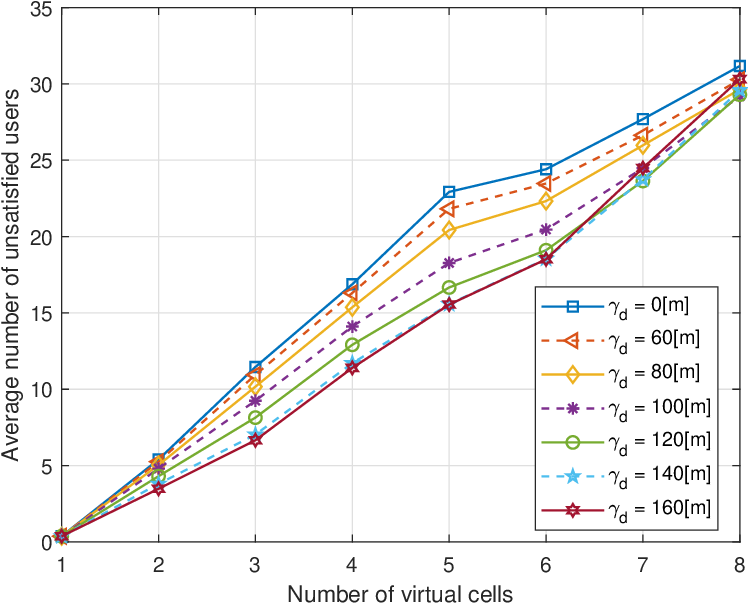}
    \caption{Number of unsatisfied users for a CGBR of 512kbps.}
    \label{fig:my_label512}
\end{figure}

Figures \ref{fig:my_label128}-\ref{fig:my_label512} depict the number of unsatisfied users as a function of the number of virtual cells in the system for CGBR of: 128kbps , 256kbps , and 512kbps , respectively. These figures show the reduction in the average number of unsatisfied users in the system that our scheme provides. We examine   several values of the threshold $\gamma_d$ between $0$ and $140$ meters where $\gamma_d=0$ is the special case with no interference reduction between virtual cells. 

\begin{figure}
    \centering
    \includegraphics[scale=0.66]{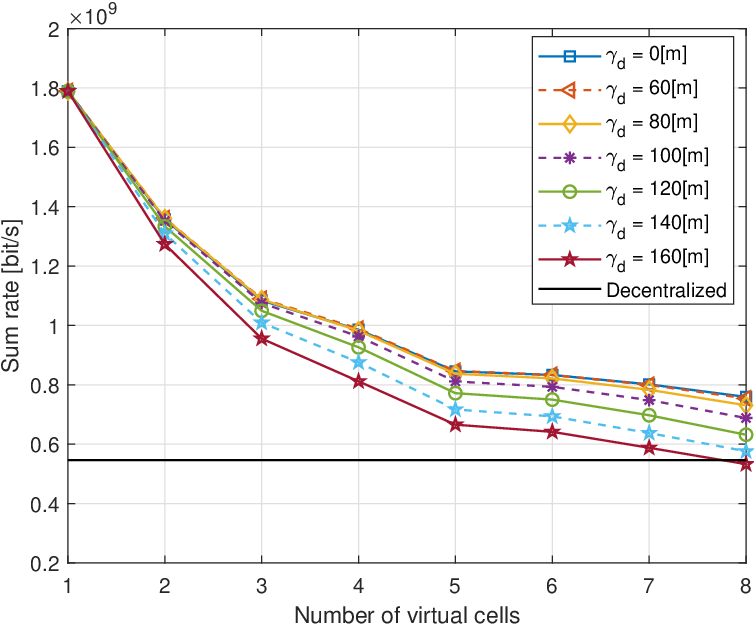}
    \caption{A comparison of system sum rate for each of several values of the threshold $\gamma_d$. The black line represents the sum rate of a decentralized network.}
    \label{fig:my_label_rate}
\end{figure}

Figures \ref{fig:my_label128}-\ref{fig:my_label512} show that as  the number of virtual cells in the system decreases, the number of unsatisfied users decreases as well. 
Furthremore, Figures \ref{fig:my_label128}-\ref{fig:my_label256}  show that as the interference distance $\gamma_d$ increases, the number of unsatisfied users decreases. As we increase the CGBR to $512$kbps  we see from Figure \ref{fig:my_label512} that increasing $\gamma_d$ beyond $140$m, may not provide a reduction in the number of unsatisfied users when the network is clustered into $8$  virtual cells. 
This occurs since increasing the number of virtual cells increases the number of interfering edge BSs and decreases the system sum rate (see Figure \ref{fig:my_label_rate}). Additionally, increasing the threshold $\gamma_d$  reduces the interference experienced in each allocated frequency band by reducing the number of frequency bands, i.e. $|\mathcal{K}_b|$, allocated to each edge BS $b$. Therefore,  when the number of virtual cells increases,  large values of $\gamma_d$  prevent users from achieving higher communication rates since the interference reduction  they provide   does not compensate for the reduction in the  number of  frequency bands  allocated to each edge BS they cause,  
this is evident in the case where there are $8$ virtual cells and $\gamma_d=140$m in  Figure \ref{fig:my_label512}.       
Overall Figures \ref{fig:my_label128}-\ref{fig:my_label512} show that both increasing the threshold $\gamma_d$ and decreasing the number of virtual cells  can significantly reduce the number of unsatisfied users in the system.  
However,  reducing the number of virtual cells increases the encoding and decoding complexity and requires additional acquisition of channel state information compared with a larger number of virtual cells. In contrast, the interference reduction scheme we propose in this paper decreases the number of unsatisfied users without the overhead that reducing the number of virtual cells in the system requires. 

To understand the impact of prioritizing the improvement of the minimal rate achieved by users, Figure \ref{fig:my_label_rate} evaluates the system sum rate of each of the considered threshold distance values $\gamma_d$. The black line depicts the system sum rate of a decentralized system.  We can conclude from  Figures \ref{fig:my_label128}-\ref{fig:my_label_rate} that increasing the values of $\gamma_d$ beyond $140$m  does not decrease the number of unsatisfied users dramatically, but reduces the system sum rate  significantly. Thus the choice of the value $\gamma_d$ should balance the requirement to minimize the number of unsatisfied users with the desire to increase the system sum rate.

\section{Conclusion}\label{sec:conclusion}

This work has addressed the problem of channel and power allocation in cellular networks based on virtual cells, whereby interference from both inside and outside the virtual cells is considered. The interference from within a virtual cell is managed by optimizing the power allocation in the virtual cell.  Interference from outside a virtual cell to its edge users is reduced  by frequency allocation. This frequency allocation is performed by creating an interference graph between BSs at the edges of interfering virtual cells where two BSs of different virtual cells are considered as interfering if their distance does not exceed a certain threshold. Since we perform cooperative decoding, the frequency allocation scheme that limit BSs receiving frequency bands is only used in the power allocation scheme. At the decoding stage the covariance matrix of the interference is known and BSs use the signals received over all frequency bands to decode their messages.
 Numerical results show that our design reduces the average number of unsatisfied users. Interestingly, both decreasing the number of virtual cells in the system and 
 increasing the distance threshold of interfering BSs reduce the number of unsatisfied users. However, the latter does not require additional channel state information but uses only  the number of users that choose each BS as its most desirable BS in terms of SNR.



\begin{thebibliography}{10}
\providecommand{\url}[1]{#1}
\csname url@samestyle\endcsname
\providecommand{\newblock}{\relax}
\providecommand{\bibinfo}[2]{#2}
\providecommand{\BIBentrySTDinterwordspacing}{\spaceskip=0pt\relax}
\providecommand{\BIBentryALTinterwordstretchfactor}{4}
\providecommand{\BIBentryALTinterwordspacing}{\spaceskip=\fontdimen2\font plus
\BIBentryALTinterwordstretchfactor\fontdimen3\font minus
  \fontdimen4\font\relax}
\providecommand{\BIBforeignlanguage}[2]{{%
\expandafter\ifx\csname l@#1\endcsname\relax
\typeout{** WARNING: IEEEtran.bst: No hyphenation pattern has been}%
\typeout{** loaded for the language `#1'. Using the pattern for}%
\typeout{** the default language instead.}%
\else
\language=\csname l@#1\endcsname
\fi
#2}}
\providecommand{\BIBdecl}{\relax}
\BIBdecl

\bibitem{4623708}
V.~Chandrasekhar, J.~G. Andrews, and A.~Gatherer, ``Femtocell networks: a
  survey,'' \emph{IEEE Commun. Mag.}, vol.~46, no.~9, pp. 59--67, Sept 2008.

\bibitem{6768783}
H.~Claussen, L.~T.~W. Ho, and L.~G. Samuel, ``An overview of the femtocell
  concept,'' \emph{Bell Labs Technical Journal}, vol.~13, no.~1, pp. 221--245,
  Spring 2008.

\bibitem{6171992}
J.~G. Andrews, H.~Claussen, M.~Dohler, S.~Rangan, and M.~C. Reed, ``Femtocells:
  Past, present, and future,'' \emph{IEEE J. Sel. Areas Commun.}, vol.~30,
  no.~3, pp. 497--508, April 2012.

\bibitem{anpalagan_bennis_vannithamby_2015}
A.~Anpalagan, M.~Bennis, and R.~Vannithamby, \emph{Design and Deployment of
  Small Cell Networks}.\hskip 1em plus 0.5em minus 0.4em\relax Cambridge
  University Press, 2015.

\bibitem{5876482}
F.~{Khan} and Z.~{Pi}, ``{mmWave} mobile broadband (mmb): Unleashing the
  3–300ghz spectrum,'' in \emph{34th IEEE Sarnoff Symposium}, May 2011, pp.
  1--6.

\bibitem{5783993}
Z.~{Pi} and F.~{Khan}, ``An introduction to millimeter-wave mobile broadband
  systems,'' \emph{IEEE Commun. Mag.}, vol.~49, no.~6, pp. 101--107, June 2011.

\bibitem{6732923}
S.~{Rangan}, T.~S. {Rappaport}, and E.~{Erkip}, ``Millimeter-wave cellular
  wireless networks: Potentials and challenges,'' \emph{Proceedings of the
  IEEE}, vol. 102, no.~3, pp. 366--385, March 2014.

\bibitem{7010535}
R.~{Baldemair}, T.~{Irnich}, K.~{Balachandran}, E.~{Dahlman}, G.~{Mildh},
  Y.~{Selén}, S.~{Parkvall}, M.~{Meyer}, and A.~{Osseiran}, ``Ultra-dense
  networks in millimeter-wave frequencies,'' \emph{IEEE Commun. Mag.}, vol.~53,
  no.~1, pp. 202--208, January 2015.

\bibitem{8653366}
C.~{Fiandrino}, H.~{Assasa}, P.~{Casari}, and J.~{Widmer}, ``Scaling
  millimeter-wave networks to dense deployments and dynamic environments,''
  \emph{Proceedings of the IEEE}, vol. 107, no.~4, pp. 732--745, April 2019.

\bibitem{7839266}
S.~Bassoy, H.~Farooq, M.~A. Imran, and A.~Imran, ``Coordinated multi-point
  clustering schemes: A survey,'' \emph{IEEE Commun. Surveys Tutorials},
  vol.~19, no.~2, pp. 743--764, Secondquarter 2017.

\bibitem{4385782}
O.~Somekh, B.~M. Zaidel, and S.~{Shamai (Shitz)}, ``Sum rate characterization
  of joint multiple cell-site processing,'' \emph{IEEE Transactions on
  Information Theory}, vol.~53, no.~12, pp. 4473--4497, 2007.

\bibitem{4729770}
O.~Simeone, O.~Somekh, H.~V. Poor, and S.~{Shamai (Shitz)}, ``Local base
  station cooperation via finite-capacity links for the uplink of linear
  cellular networks,'' \emph{IEEE Transactions on Information Theory}, vol.~55,
  no.~1, pp. 190--204, 2009.

\bibitem{5351392}
O.~Somekh, O.~Simeone, H.~V. Poor, and S.~Shamai, ``Throughput of cellular
  uplink with dynamic user activity and cooperative base-stations,'' in
  \emph{2009 IEEE Information Theory Workshop}, 2009, pp. 610--614.

\bibitem{CIT-048}
\BIBentryALTinterwordspacing
O.~Simeone, N.~Levy, A.~Sanderovich, O.~Somekh, B.~M. Zaidel, H.~V. Poor, and
  S.~S. (Shitz), ``Cooperative wireless cellular systems: An
  information-theoretic view,'' \emph{Foundations and Trends® in
  Communications and Information Theory}, vol.~8, no. 1-2, pp. 1--177, 2012.
  [Online]. Available: \url{http://dx.doi.org/10.1561/0100000048}
\BIBentrySTDinterwordspacing

\bibitem{6786390}
V.~Garcia, Y.~Zhou, and J.~Shi, ``Coordinated multipoint transmission in dense
  cellular networks with user-centric adaptive clustering,'' \emph{IEEE Trans.
  Wireless Commun.}, vol.~13, no.~8, pp. 4297--4308, Aug 2014.

\bibitem{7248710}
L.~{Liu}, V.~{Garcia}, L.~{Tian}, Z.~{Pan}, and J.~{Shi}, ``Joint clustering
  and inter-cell resource allocation for {CoMP} in ultra dense cellular
  networks,'' in \emph{2015 IEEE Int. Conf. on Commun. (ICC)}, June 2015, pp.
  2560--2564.

\bibitem{8110665}
L.~{Liu}, Y.~{Zhou}, V.~{Garcia}, L.~{Tian}, and J.~{Shi}, ``Load aware joint
  {CoMP} clustering and inter-cell resource scheduling in heterogeneous ultra
  dense cellular networks,'' \emph{IEEE Trans. Veh. Technol.}, vol.~67, no.~3,
  pp. 2741--2755, March 2018.

\bibitem{7470561}
S.~Bassoy, M.~Jaber, M.~A. Imran, and P.~Xiao, ``Load aware self-organising
  user-centric dynamic {CoMP} clustering for 5g networks,'' \emph{IEEE Access},
  vol.~4, pp. 2895--2906, 2016.

\bibitem{8755921}
S.~{Bassoy}, M.~A. {Imran}, S.~{Yang}, and R.~{Tafazolli}, ``A load-aware
  clustering model for coordinated transmission in future wireless networks,''
  \emph{IEEE Access}, vol.~7, pp. 92\,693--92\,708, 2019.

\bibitem{340450}
A.~D. Wyner, ``Shannon-theoretic approach to a {Gaussian} cellular
  multiple-access channel,'' \emph{IEEE Trans. Inf. Theory}, vol.~40, no.~6,
  pp. 1713--1727, Nov 1994.

\bibitem{5371570}
J.~{Liu} and D.~{Wang}, ``An improved dynamic clustering algorithm for
  multi-user distributed antenna system,'' in \emph{2009 International
  Conference on Wireless Communications Signal Processing}, 2009, pp. 1--5.

\bibitem{4533793}
A.~Papadogiannis, D.~Gesbert, and E.~Hardouin, ``A dynamic clustering approach
  in wireless networks with multi-cell cooperative processing,'' in \emph{2008
  IEEE Int. Conf. on Commun.}, May 2008, pp. 4033--4037.

\bibitem{6398781}
H.~{Li}, H.~{Tian}, C.~{Qin}, and Y.~{Pei}, ``A novel distributed cluster
  combination method for comp in lte-a system,'' in \emph{The 15th
  International Symposium on Wireless Personal Multimedia Communications},
  2012, pp. 614--618.

\bibitem{6488494}
{Jung-Min Moon} and {Dong-Ho Cho}, ``Formation of cooperative cluster for
  coordinated transmission in multi-cell wireless networks,'' in \emph{2013
  IEEE 10th Consumer Communications and Networking Conference (CCNC)}, 2013,
  pp. 528--533.

\bibitem{6858023}
F.~{Guidolin}, L.~{Badia}, and M.~{Zorzi}, ``A distributed clustering algorithm
  for coordinated multipoint in {LTE} networks,'' \emph{IEEE Wireless
  Communications Letters}, vol.~3, no.~5, pp. 517--520, 2014.

\bibitem{MarschICC2011}
P.~Marsch and G.~Fettweis, ``Static clustering for cooperative multipoint
  {(CoMP)} in mobile communications,'' in \emph{Proceedings of IEEE
  International Conference on Communications (ICC)}, 2011, pp. 1--6.

\bibitem{6530435}
S.~S. Ali and N.~Saxena, ``A novel static clustering approach for {CoMP},'' in
  \emph{2012 7th Int. Conf. on Computing and Convergence Technology (ICCCT)},
  Dec 2012, pp. 757--762.

\bibitem{6707857}
H.~Shimodaira, G.~K. Tran, K.~Araki, K.~Sakaguchi, S.~Konishi, and S.~Nanba,
  ``Diamond cellular network — optimal combination of small power
  basestations and {CoMP} cellular networks-,'' in \emph{2013 IEEE 24th Int.
  Symposium on Personal, Indoor and Mobile Radio Commun. (PIMRC Workshops)},
  Sept 2013, pp. 163--167.

\bibitem{9343767}
S.~Gelincik, M.~Wigger, and L.~Wang, ``Benefits of local cooperation in
  sectorized cellular networks under a complexity constraint,'' \emph{IEEE
  Transactions on Wireless Communications}, vol.~20, no.~6, pp. 3897--3910,
  2021.

\bibitem{YeminiGoldsmith1}
M.~Yemini and A.~J. Goldsmith, ``Virtual cell clustering with optimal resource
  allocation to maximize cellular system capacity,'' in \emph{2019 IEEE Global
  Communications Conference (Globecom)}, December 2019.

\bibitem{YeminiGoldsmith2}
------, ``Optimal resource allocation for cellular networks with virtual cell
  joint decoding,'' in \emph{2019 IEEE Int. Symposium on Inf. Theory (ISIT)},
  July 2019.

\bibitem{YeminiGoldsmithJounal}
M.~{Yemini} and A.~J. {Goldsmith}, ``Virtual cell clustering with optimal
  resource allocation to maximize capacity,'' \emph{IEEE Transactions on
  Wireless Communications}, vol. Early access, March 2021.

\bibitem{5594708}
D.~Gesbert, S.~Hanly, H.~Huang, S.~{Shamai (Shitz)}, O.~Simeone, and W.~Yu,
  ``Multi-cell mimo cooperative networks: A new look at interference,''
  \emph{IEEE Journal on Selected Areas in Communications}, vol.~28, no.~9, pp.
  1380--1408, 2010.

\bibitem{Bertsekas/99}
D.~Bertsekas, \emph{Nonlinear Programming}.\hskip 1em plus 0.5em minus
  0.4em\relax Athena Scientific, 1999.

\bibitem{1262622}
{Wei Yu}, {Wonjong Rhee}, S.~{Boyd}, and J.~M. {Cioffi}, ``Iterative
  water-filling for gaussian vector multiple-access channels,'' \emph{IEEE
  Transactions on Information Theory}, vol.~50, no.~1, pp. 145--152, Jan 2004.

\bibitem{BienTibshirani2011}
J.~Bien and R.~Tibshirani, ``Hierarchical clustering with prototypes via
  minimax linkage,'' \emph{Journal of the American Statistical Association},
  vol. 106, no. 495, pp. 1075--1084, 2011.

\bibitem{RLF}
F.~T. Leighton, ``A graph coloring algorithm for large scheduling problems,''
  \emph{JOURNAL OF RESEARCH of the Notional Bureau of Standards}, vol.~84,
  no.~6, pp. 489--506, Nov.-Dec. 1979.

\bibitem{Adegbindin2014ANE}
M.~Adegbindin, A.~Hertz, and M.~Bellaiche, ``A new efficient {RLF}-like
  algorithm for the vertex coloring problem,'' \emph{Yugoslav Journal of
  Operations Research}, vol.~26, pp. 441--456, 2014.

\bibitem{6834753}
M.~R. {Akdeniz}, Y.~{Liu}, M.~K. {Samimi}, S.~{Sun}, S.~{Rangan}, T.~S.
  {Rappaport}, and E.~{Erkip}, ``Millimeter wave channel modeling and cellular
  capacity evaluation,'' \emph{IEEE Journal on Selected Areas in
  Communications}, vol.~32, no.~6, pp. 1164--1179, June 2014.

\end{thebibliography}

\balance
\bibliographystyle{IEEEtran}

\end{document}